\documentclass[doublecol]{epl2}

\usepackage{amsmath,mathrsfs}
\usepackage[dvipsnames]{xcolor}

\newcommand{\prt}{\partial}

\newcommand{\ga}{\gamma}
\newcommand{\sss}{\scriptscriptstyle}

\newcommand{\ox}{\overline{x}}

\title{Dispersionless evolution of inviscid nonlinear pulses}

\author{M. Isoard\inst{1} \and A. M. Kamchatnov\inst{2,3} \and
  N. Pavloff\inst{1}} \shortauthor{M. Isoard \etal}

\institute{
  \inst{1} Universit\'e Paris-Saclay, CNRS,
  LPTMS, 91405 Orsay, France \\
  \inst{2} Institute of Spectroscopy,
  Russian Academy of Sciences, Troitsk, Moscow, 108840, Russia\\
  \inst{3} Moscow Institute of Physics and Technology, Institutsky
  lane 9, Dolgoprudny, Moscow region, 141701, Russia
}


\abstract{We consider the one-dimensional dynamics of nonlinear
  non-dispersive waves. The problem can be mapped onto a linear one by
  means of the hodograph transform. We propose an approximate scheme
  for solving the corresponding Euler-Poisson equation which is valid
  for any kind of nonlinearity. The approach is exact for monoatomic
  classical gas and agrees very well with exact results and numerical
  simulations for other systems. We also provide a simple and accurate
  determination of the wave breaking time for typical initial
  conditions.}

\begin{document}

\maketitle

\section{Introduction}

In the long wavelength limit, many physical models lead, in the
one-dimensional regime, to equations of wave propagation equivalent to
the equations of inviscid gas dynamics
\begin{equation}\label{1-1}
  \rho_t+(\rho u)_x=0,\qquad u_t+uu_x+\frac{c^2}{\rho}\rho_x=0,
\end{equation}
where $u$ is interpreted as a local ``flow velocity'', and
  $c=c(\rho)$ has a meaning of a local ``sound velocity'' which
  depends on a local ``density'' $\rho$. These nonlinear equations
were studied very intensively in the framework of gas dynamics (see,
e.g., Ref.~\cite{LL-6}) and a number of exact solutions have been
obtained for various problems in the particular case of polytropic
gases for which, up to a normalization constant which can be rescaled
to unity:
\begin{equation}\label{1-2a}
  c(\rho)= \rho^{(\ga-1)/2},
\end{equation}
where $\gamma$ is the adiabatic index ($\gamma>1$).  However, even in
this apparently simple case, the solutions become quite complicated if
the parameter
\begin{equation}\label{1-2}
  \beta=\frac{3-\ga}{2(\ga-1)}
\end{equation}
is not an integer number. This difficulty is encountered for instance
in the study of the evolution of a nonlinear pulse with initial
density and velocity distributions
\begin{equation}\label{1-3}
  \rho(x,0)=\overline{\rho}(x), \quad u(x,0)=0,
\end{equation}
where $\overline{\rho}$ is a specified function of $x$; see, e.g., the
solution of the problems pulse evolution in optical systems with Kerr
nonlinearity \cite{For2009} or of collision of two rarefaction waves in
the dynamics of a Bose-Einstein condensed system \cite{ik-19}, for which
$\ga=2$ (and $\beta=1/2$).

In Refs.~\cite{ikp-19a,ikp-19b} it was noticed that for the case
$\ga=2$ one can obtain a very accurate and simple {\it approximate}
solution of the problem of evolution of the pulse (\ref{1-3}). The aim
of the present paper is to generalize this approach to arbitrary dependence
$c=c(\rho)$. We first present the hodograph transform which maps the
nonlinear system onto a linear Euler-Poisson equation which can be
solved by Riemann's method. We then propose an approximate expression
for the Riemann function which leads to a simple solution of the
problem. The approach is discussed and compared with numerical
simulations. We also discuss an approximate determination of the time
of shock formation in the system.

\section{Hodograph transform and Riemann method}

The term in $\rho_x$ in (\ref{1-1}) being positive, the system is
hyperbolic. It can be cast to a diagonal form by introducing the
Riemann invariants
\begin{equation}\label{2-5}
r_{\pm}(x,t)=\tfrac12 u(x,t)\pm
  \tfrac12 \int_0^{\rho(x,t)}\frac{c(\rho')\upd\rho'}{\rho'}.
\end{equation}
$r_+$ and $r_-$ obey dynamical equations equivalent to (\ref{1-1}) which
  take the form
\begin{equation}\label{2-6}
  \frac{\prt r_{\pm}}{\prt t}+v_{\pm}\frac{\prt r_{\pm}}{\prt x}=0,
\end{equation}
where
\begin{equation}\label{2-7}
  v_{\pm}=u\pm c
\end{equation}
can be expressed in terms of the Riemann invariants. Indeed, it
follows from Eq.~(\ref{2-5}) that the physical variables $u$ and $c$
can be written as
\begin{equation}\label{2-8}
  u=r_++r_- ,\qquad c=c(r_+-r_-),
\end{equation}
where the expression of $c$ as a function of $r_+-r_-$ is obtained by
inverting the relation\footnote{The dependence of $c$ on
  $r_+-r_-$ is different from its dependence on $\rho$. In the
  following, we always specify the argument of $c$ to avoid
  confusion.}
\begin{equation}\label{rderho}
  r_+-r_-=\int_0^{\rho}\frac{c(\rho')\upd\rho'}{\rho'}
\end{equation}
and substituting $\rho(r_+-r_-)$ into $c=c(\rho)$.
Then, the velocities $v_\pm$ in Eq.~(\ref{2-7}) can be considered as
known functions of $r_+$ and $r_-$:
\begin{equation}\label{2-9}
  v_{\pm}(r_+,r_-)=r_++r_- \pm c(r_+-r_-).
\end{equation}
The equations (\ref{2-6}) can be linearized by the hodograph
transform (see, e.g., Refs.~\cite{LL-6,kamch-2000}).
This consists in considering $x$ and $t$ as functions of
the independent variables $r_{\pm}$ and leads to the following
system of linear equations:
\begin{equation}\label{2-10}
\begin{split}
&  \frac{\prt x}{\prt r_+}-v_-(r_+,r_-)\frac{\prt t}{\prt r_+}=0,\\
&  \frac{\prt x}{\prt r_-}-v_+(r_+,r_-)\frac{\prt t}{\prt r_-}=0.
\end{split}
\end{equation}
We look for the solutions of these equations in the form
\begin{equation}\label{2-11}
\begin{split}
&    x-v_+(r_+,r_-)t=w_+(r_+,r_-),\\
& x-v_-(r_+,r_-)t=w_-(r_+,r_-).
\end{split}
\end{equation}
A simple test of consistency shows that the unknown functions $w_\pm(r_+,r_-)$
should verify the Tsarev equations \cite{Tsa91}
\begin{equation}\label{2-12}
\begin{split}
& \frac1{w_+-w_-}\frac{\prt w_+}{\prt r_-}=
\frac1{v_+-v_-}\frac{\prt v_+}{\prt r_-},\\
& \frac1{w_+-w_-}\frac{\prt w_-}{\prt r_+}=
\frac1{v_+-v_-}\frac{\prt v_-}{\prt r_+}.
\end{split}
\end{equation}
Now we notice that since the velocities $v_{\pm}$ are given by expressions
(\ref{2-9}), the
right-hand sides of both Eqs.~(\ref{2-12}) are equal to each other:
\begin{equation}\label{3-13}
  \frac1{v_+-v_-}\frac{\prt v_+}{\prt r_-}=
\frac1{v_+-v_-}\frac{\prt v_-}{\prt r_+}=
  \frac{1-c'(r_+-r_-)}{2\, c(r_+-r_-)},
\end{equation}
where $c'(r)\equiv \upd c(r)/\upd r$. Consequently $\prt w_+/\prt
r_-=\prt w_-/\prt r_+$ and $w_{\pm}$ can be sought in the form
\begin{equation}\label{3-15}
  w_{\pm}=\frac{\prt W}{\prt r_{\pm}}.
\end{equation}
Substitution of Eqs.~(\ref{3-13}) and (\ref{3-15}) into
Eqs.~(\ref{2-12}) shows that the function $W$ obeys the Euler-Poisson
equation
\begin{equation}\label{3-16}
  \frac{\prt^2 W}{\prt r_+\prt r_-}
  -\frac{1-c'(r_+-r_-)}{2c(r_+-r_-)}
  \left(\frac{\prt W}{\prt r_+}-\frac{\prt W}{\prt r_-}\right)=0.
\end{equation}
A formal solution of Eq.~(\ref{3-16}) in the $(r_+,r_-)$ plane (the
so-called hodograph plane) can be obtained with the use of the Riemann
method (see, e.g., Ref.~\cite{Som64}). We introduce the notation
\begin{equation}
a(r_+,r_-)=\frac{c'(r_+-r_-)-1}{2c(r_+-r_-)}=-b(r_+,r_-),
\end{equation}
and the so-called Riemann function $R(r_+,r_-;\xi,\eta)$
which satisfies an equation conjugate to \eqref{3-16}
\begin{equation}\label{3-19}
  \frac{\prt^2 R}{\prt r_+\prt r_-}-\frac{\prt (a R)}{\prt r_+}
-\frac{\prt (b R)}{\prt r_-}=0,
\end{equation}
with the boundary conditions:
\begin{equation}\label{3-20}
  \begin{split}
  & \frac{\prt R}{\prt r_+}-bR=0\quad
\text{along the characteristic}\quad r_-=\eta,\\
  & \frac{\prt R}{\prt r_-}-aR=0\quad
\text{along the characteristic}\quad r_+=\xi,
  \end{split}
\end{equation}
and
\begin{equation}\label{3-21}
  R(\xi,\eta;\xi,\eta)=1.
\end{equation}
Then, at a point $P$ with coordinates $(\xi,\eta)$ of
the hodograph plane, $W$ can be expressed as:
\begin{equation}\label{3-18}
  W(P)=\tfrac12\left(R W\right)_{\! A}
  + \tfrac12\left(R W\right)_{\! B}
  - \int_A^B\!\!(V\upd r_++U\upd r_-),
\end{equation}
where
\begin{equation}\label{3-17}
  \begin{split}
  & U=\frac12
\left(R \frac{\prt W}{\prt r_-}-
W\, \frac{\prt R}{\prt r_-}\right)
  +aR W,\\
&  V=\frac12\left( W\,\frac{\prt R}{\prt r_+}-
R\frac{\prt W}{\prt r_+}\right)
  -bRW.
  \end{split}
\end{equation}
We use here doubled notation for the coordinates in the hodograph
plane: $(\xi,\eta)$ and $(r_+,r_-)$. $P=(\xi,\eta)$ is the
``observation'' point and the integral in (\ref{3-18}) is taken over
the curve $\mathcal{C}$ of the initial data in this plane which has
parametric equation $(r_+(x,0),r_-(x,0))$. The points $A$ and $B$ are
projections of $P$ onto $\mathcal{C}$ along the $r_+$ and $r_-$ axis
respectively.  The advantage of the expression (\ref{3-18}) is that it
gives the value of $W$ at $P$ in terms of its values (and of the one
of its derivatives) along the curve $\mathcal{C}$ of initial
conditions.

Once the Riemann function $R$ has been determined, Eq.~(\ref{3-18})
gives the solution of the problem under consideration.

\section{Approximate solution}

We now proceed and consider the specific problem formulated in the
Introduction. To simplify the discussion we assume that the initial
distribution $\overline{\rho}(x)$ reaches an extremum at $x=0$ and is
an even function of $x$: $\overline{\rho}(-x)=\overline{\rho}(x)$. The
generalization to non-symmetric distributions is straightforward.

First of all, we have to understand how the initial profile
(\ref{1-3}) fixes the boundary conditions for $W$ on curve
$\mathcal{C}$ in the hodograph plane. To this end, we compute the
initial distribution of the Riemann invariant $r_+$ for positive $x$
at $t=0$:
\begin{equation}\label{4-22}
\overline{r}(x)\equiv
r_+(x>0,t=0)=\frac{1}{2}
\int_0^{\overline{\rho}(x)}\frac{c(\rho)\upd\rho}{\rho}.
\end{equation}
Denoting as $\ox(r)$ the reciprocal function, we
obtain for the value of $x$ on the curve $\mathcal{C}$:
\begin{equation}\label{4-23}
  x=\begin{cases}\phantom{-}\ox(r) & \mbox{if}\; x>0,\\
-\ox(r) & \mbox{if}\; x<0.\end{cases}
\end{equation}
Besides that, since at $t=0$ the values $r_+=\overline{r}$ and
$r_-=-\overline{r}$ correspond to the same value of $x$, we find that
$\ox(r)$ is an even function, $\ox(-r)=\ox(r)$. As an illustration,
for an initial profile of the form
\begin{equation}\label{rho-init}
\overline{\rho}(x)=\rho_0+\rho_1\exp(-x^2/x_0^2)\; ,
\end{equation}
(with $\rho_0$ and $\rho_1>0$) one obtains in the case of a polytropic
gas \eqref{1-2a}:
\begin{equation}
\ox(r)=x_0\sqrt{\ln\rho_1-\ln\left[
\left|(\gamma-1)\, r\right|^{\frac{2}{\gamma-1}}-\rho_0\right]}.
\end{equation}

\begin{figure}
\centering
\includegraphics[width=\linewidth]{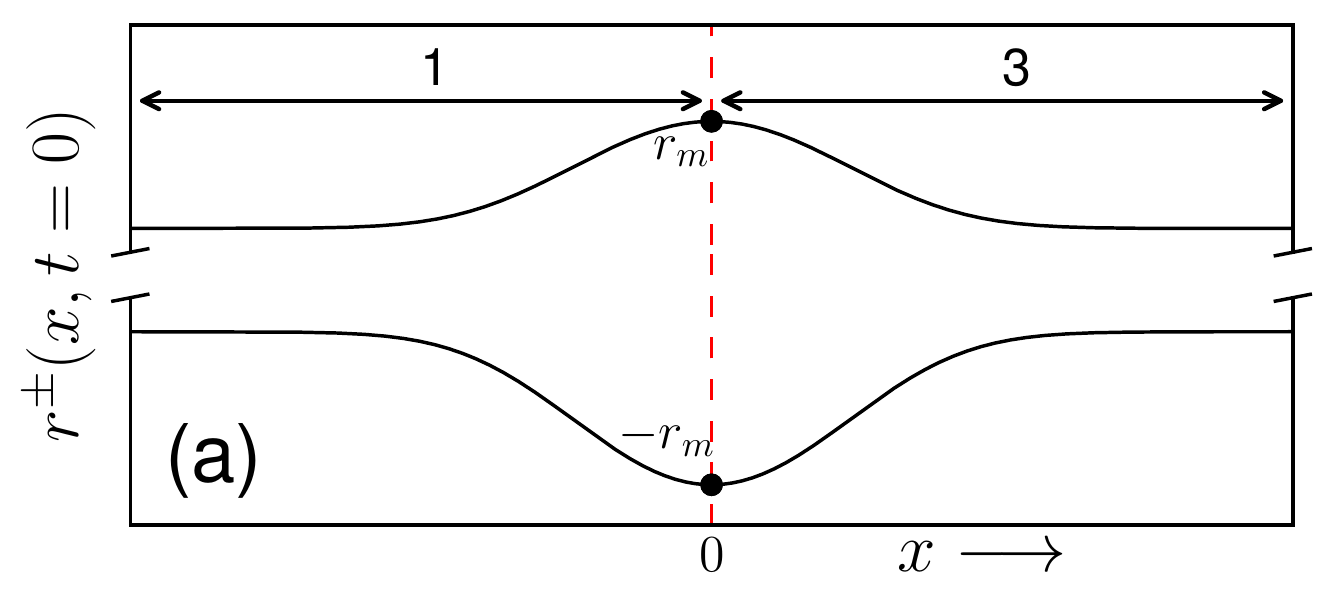}
\includegraphics[width=\linewidth]{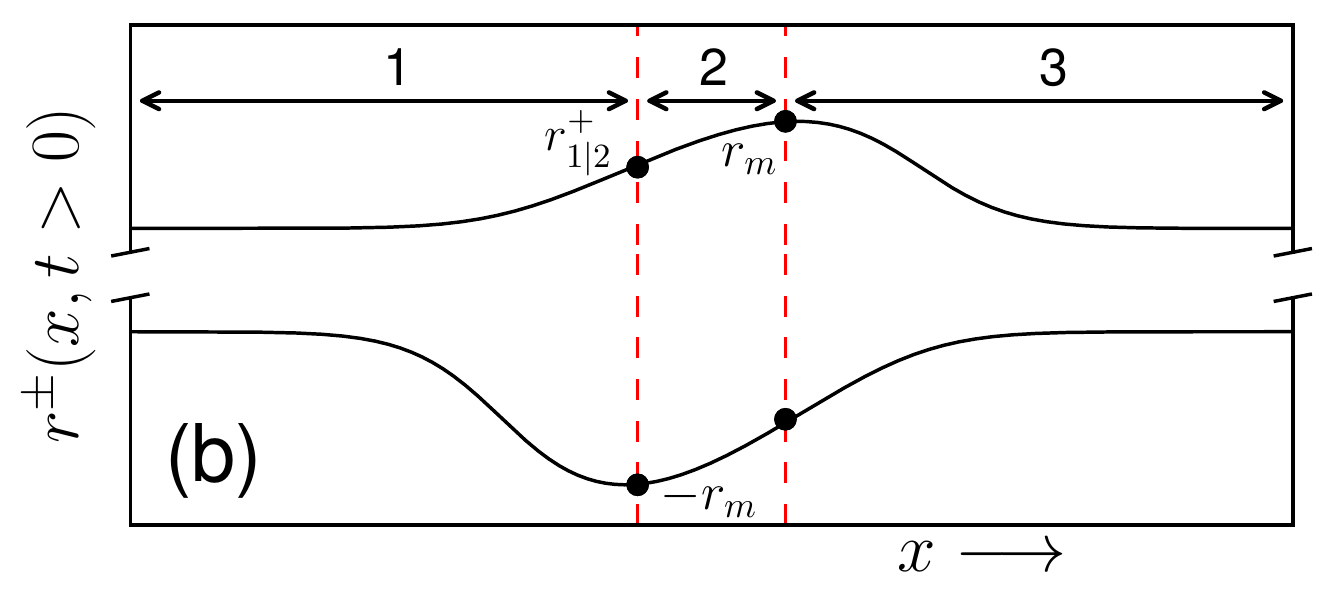}
\caption{Sketch of the distributions $r^\pm(x,t)$ at time $t=0$ (a)
  and at finite time $t>0$ (b). In each panel the upper solid curve
  represents $r^+$ (always larger than $r_0$), and the lower one $r^-$
  (always lower than $-r_0$). For $t>0$, $r^+$ ($r^-$) moves to the
  right (to the left) and regions 1 and 3 start to overlap. This
  leads to the configuration represented in panel (b) where a new
  region (labeled region 2) has appeared. The value of $r^+$ at the
  interface between regions 1 and 2 is denoted as $r_{1|2}^+$.}
\label{fig1}
\end{figure}

\begin{figure}
\begin{picture}(8.,14.5)
\put(0.4,0){\includegraphics[width=8.cm]{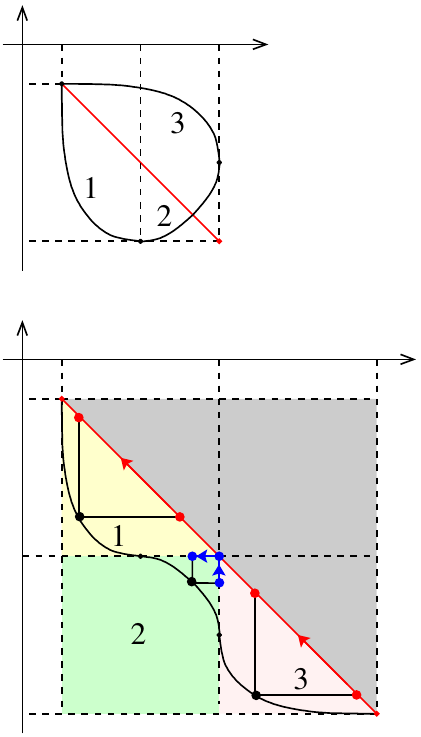}}
\put(5.4,12){\huge(a)}
\put(5.4,8.5){\huge(b)}
\put(2.6,11.5){\LARGE\color{red}${\cal C}$}
\put(2.8,5.5){\LARGE\color{red}${\cal C}$}
\put(0.5,13.9){\large$r^-$}
\put(5.3,13.4){\large$r^+$}
\put(1.4,13.4){\large$r_0$}
\put(2.8,13.4){\large$r_{1|2}^+$}
\put(4.3,13.4){\large$r_m$}
\put(0,12.4){\large$-r_0$}
\put(0,9.4){\large$-r_m$}
\put(0.5,7.9){\large$r^-$}
\put(7.95,7.4){\large$r^+$}
\put(2.1,7.5){\small$r^+$ increases}
\put(2.1,7.85){\vector(1,0){1.8}}
\put(5.1,7.5){\small$r^+$ decreases}
\put(5.1,7.85){\vector(1,0){1.8}}
\put(1.4,7.4){\large$r_0$}
\put(4.3,7.4){\large$r_m$}
\put(7.35,7.4){\large$r_0$}
\put(0.2,6.0){\rotatebox{-90}{\small$r^-$ decreases}}
\put(0.6,6.0){\vector(0,-1){1.8}}
\put(0.2,2.9){\rotatebox{-90}{\small$r^-$ increases}}
\put(0.6,2.9){\vector(0,-1){1.8}}
\put(0,6.4){\large$-r_0$}
\put(0,3.4){\large$-r_m$}
\put(0,0.4){\large$-r_0$}
\put(2,4.4){\large$P$}
\put(2.05,6.0){\large\color{red}$B_1$}
\put(3.9,4.5){\large\color{red}$A_1$}
\put(3.65,2.6){\large$P$}
\put(3.7,3.6){\large\color{blue}$B_2$}
\put(4.6,2.7){\large\color{blue}$A_2$}
\put(4.65,3.6){\large\color{blue}$C$}
\put(5.3,1.0){\large$P$}
\put(5.3,2.8){\large\color{red}$B_3$}
\put(7.0,1.1){\large\color{red}$A_3$}
\end{picture}
\caption{(a) Behavior of the Riemann invariants in the characteristic
  plane at a given time $t$. The red straight line is the curve
  $\mathcal{C}$. The black solid line is the curve with parametric
  equation $(r_+(x,t),r_-(x,t))$. (b) The same curves on the
  four-sheeted unfolded surface. The colored regions 1, 2 and 3 are
  the same as the ones identified in Fig.~\ref{fig1}.  In our problem,
  the whole gray shaded domain above ${\cal C}$ is unreachable. A
  generic point $P$ of has coordinates ($\xi,\eta$) and points $A_1$,
  $B_1$, $A_3$, $B_3$ and $C$ lie on the initial curve
  $\mathcal{C}$. Points $A_2$ and $B_2$ lie on a boundary between two
  regions. The arrows indicate the direction of integration in
  Eqs.~\eqref{3-18} and \eqref{5-30}.}\label{fig2}
\end{figure}
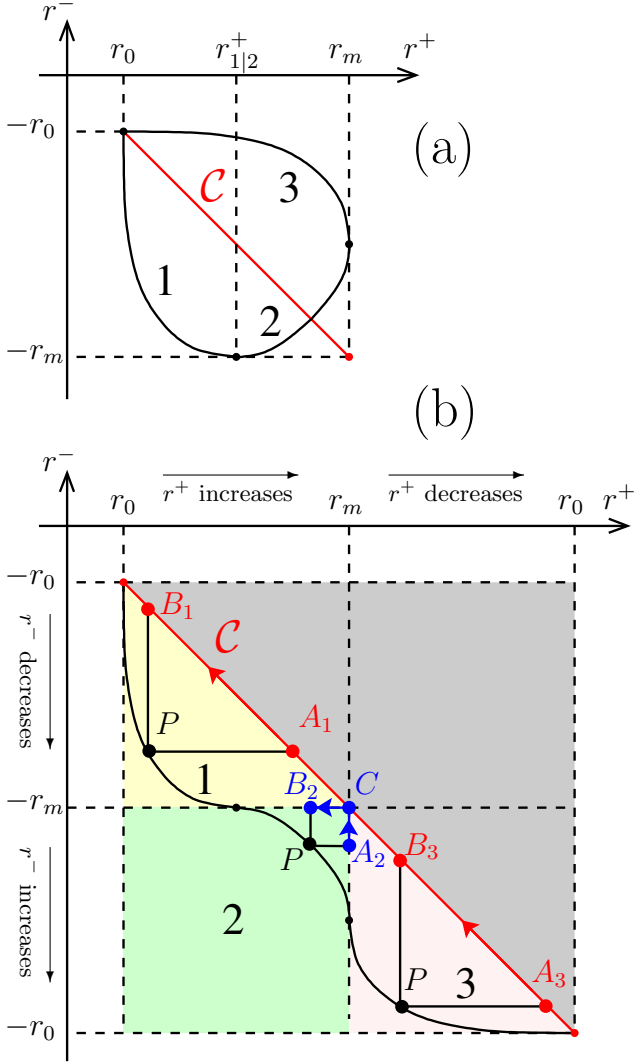

For such a ``single-bump'' type of initial conditions which we consider,
there exist two values $r_m=\overline{r}(0)$ and
$r_0=\overline{r}(x\to\infty)$ ($r_m>r_0>0$) such that
$r_+(x,t)\in [r_0,r_m]$ and $r_-(x,t)\in[-r_m,-r_0]$. At a given time
$t$, the space can be separated in three different regions 1, 2 and
3, depending on the values of $\prt r_\pm/\prt x$, as illustrated in
Fig.~\ref{fig1}. In each region both $r_+$ and $r_-$ vary
concomitantly, and this is the reason why we have to resort to
Riemann's method\footnote{We note that for some initial conditions
  there might also exist simple-wave regions which cannot be tackled
  by the Riemann method. The density and velocity profiles in such
  regions are easily described (see the case studied in
  Ref. \cite{ikp-19a}) and we do not consider here their possible
  occurrence so as not to burden the discussion.}.
For determining the value of $W$ in each of the three regions
1, 2 and 3, we follow Ludford~\cite{Lud52} and unfold the
hodograph plane into three sheets as illustrated in
Fig.~\ref{fig2}(b).

For the specific initial condition (\ref{1-3}), the curve
$\mathcal{C}$ is represented by the anti-diagonal ($r_-=-r_+$) and the
points $A$ and $B$ of Eq.~\eqref{3-18} have coordinates
$A(-\eta,\eta)$ and $B(\xi,-\xi)$. Eqs.~(\ref{2-11}) with $t=0$ give
\begin{equation*}
\left.\frac{\prt W}{\prt r_+}\right|_{\mathcal{C}}=
\left.\frac{\prt W}{\prt r_-}\right|_{\mathcal{C}}=x .
\end{equation*}
This implies that $W$ keeps a constant value along
$\mathcal{C}$. The value of this constant is immaterial, we take
$W|_{\mathcal{C}}=0$ for simplicity, and Eqs.~(\ref{3-17}) then reduce
to
\begin{equation}\label{4-25}
  U=\frac{x}{2}R(r,-r;\xi,\eta),\;\; V=-\frac{x}{2}R(r,-r;\xi,\eta).
\end{equation}
We thus obtain from (\ref{3-18}) $W=\int_{-\eta}^\xi xR \upd r$ which
gives in regions 1 and 3 the explicit expressions
\begin{equation}\label{5-26}
  W^{(1,3)}(\xi,\eta)
  =\mp\int_{-\eta}^{\xi}\ox(r)R(r,-r;\xi,\eta)\upd r,
\end{equation}
where the sign $-$ ($+$) applies in region 1 (3). The difference in
signs comes from the fact that $x=\mp\,\ox(r)$
depending on if one is in region 1 or 3 [see Eq. (\ref{4-23})].

When $P$ is in region 2 one applies formula (\ref{3-18})
with an integration path different from the one used in regions
1 and 3, see Fig.~\ref{fig2}. Upon integrating by parts one
obtains
\begin{equation}\label{5-30}
\begin{split}
  W^{(2)}(P)=& \left(R W^{(1)}\right)_{B_2} +
\left(R W^{(3)}\right)_{A_2}\\
& +
  \int_{A_2}^C\left(\frac{\prt R}{\prt r_-}-aR\right)_{r_+=r_m}
\!\!\!\!\!\!\!\!\! W^{(3)}\, \upd r_-\\
& -
  \int_C^{B_2}\left(\frac{\prt R}{\prt r_+}-bR\right)_{r_-=-r_m}
\!\!\!\!\!\!\!\!\! W^{(1)}\, \upd r_+,
\end{split}
\end{equation}
where the coordinates of the relevant points are: $A_2(r_m,\eta)$,
$B_2(\xi,-r_m)$ and $C(r_m,-r_m)$ (see Fig. \ref{fig2}).  For small
enough time of evolution, $\xi$ is close to $r_m$ and $\eta$ is close
to $-r_m$, the integrand functions in Eq.~(\ref{5-30}) are then small
by virtue of Eqs.~(\ref{3-20}). A simple
approximation thus consists in keeping only the two first terms in the
right-hand side of (\ref{5-30}).

It now remains to determine the Riemann function $R$ for computing
expression \eqref{5-26} of $W$ in regions 1 and 3 and completely
solving the problem. One first remarks that the conditions
(\ref{3-20}) and (\ref{3-21}) yield
\begin{equation}\label{5-28}
  \begin{split}
&  R(r_+,\eta;\xi,\eta)=\sqrt{\frac{c(\xi-\eta)}{c(r_+-\eta)}}
    \exp\left(\int_{\xi-\eta}^{r_+-\eta}\!\!\!
    \frac{\upd r}{2\, c(r)}\right), \\
&  R(\xi,r_-;\xi,\eta)=\sqrt{\frac{c(\xi-\eta)}{c(\xi-r_-)}}
    \exp\left(\int_{\xi-\eta}^{\xi-r_-}\!\!\!
    \frac{\upd r}{2\, c(r)}\right).
  \end{split}
\end{equation}
These expressions
suggest that $R$ can be sought in the form
\begin{equation}\label{5-29a}
  R(r_+,r_-;\xi,\eta)={\mathscr R}(r_+-r_-,\xi-\eta) F(r_+,r_-;\xi,\eta),
\end{equation}
where $F(r_+,\eta;\xi,\eta)=1=F(\xi,r_-;\xi,\eta)$ and
\begin{equation}\label{Rapprox}
\begin{split}
  {\mathscr R}(r_1,r_2)& =\sqrt{\frac{c(r_2)}{c(r_1)}}\,
  \exp\left(\int_{r_2}^{r_1}\!\frac{\upd r}{2\, c(r)}\right)\\
  & =\sqrt{\frac{c(r_2) \rho(r_1)}{c(r_1) \rho(r_2)}}.
  \end{split}
\end{equation}
The final expression in the above formula has been obtained
by means of a change of variable $\rho=\rho(r)$ in the integral, where
the function $\rho(r)$ is the reciprocal function of $r(\rho)$ given
in \eqref{rderho}
\begin{equation}\label{rderhobis}
  r(\rho)=\int_0^\rho \frac{c(\rho')\upd\rho'}{\rho'},
  \end{equation}
and $c(r)=c(\rho(r))$, so that $\upd r/c(r)=\upd \rho/\rho$.

We note here that
the approximation previously used for discarding the integrated terms in
the right-hand side of Eq.~(\ref{5-30}) amounts to assume that
$F\simeq 1$. Similarly, in expression (\ref{5-26}) for $W^{(1)}$ and
$W^{(3)}$, since at short time $-\eta$ and $\xi$ are close, the
integration variable $r$ is close to $\xi$ and one can again assume
that $F\simeq 1$. That is to say, we are led to make in the whole
hodograph plane the approximation
\begin{equation}\label{Rapprox2}
  R(r_+,r_-;\xi,\eta)\simeq {\mathscr R}(r_+-r_-,\xi-\eta).
\end{equation}
We can now write the final approximate results, making
the replacements $\xi\to r_+$, $\eta\to r_-$ in the above
expressions, so that they can be used in Eqs.~(\ref{2-11}) and
(\ref{3-15}):
\begin{equation}\label{final}
\begin{split}
& W^{(1,3)}(r_+,r_-)\simeq \mp\int_{-r_-}^{r^+}\!\!\!\!\ox(r)\,
{\mathscr R}(2 r,r_+-r_-)\upd r,\\
& W^{(2)}(r_+,r_-)\simeq
\, {\mathscr R}(r_++r_m,r_+-r_-)\, W^{(1)}(r_+,-r_m)\\
& \phantom{W^{2}(r_+,r_-)}
+ {\mathscr R}(r_m-r_-,r_+-r_-)\, W^{(3)}(r_m,r_-),
\end{split}
\end{equation}
where ${\mathscr R}$ is given by Eq.~(\ref{Rapprox}).
Formulae \eqref{Rapprox2} and (\ref{final})
are the main results of the present work. It is important to
stress that Eq.~(\ref{Rapprox}) has a universal form and can be
applied to any physical system with known dependence $c(r_+-r_-)$,
see Eqs.~(\ref{2-8}) and (\ref{rderho}).

\section{Examples}

In the case of the dynamics of a polytropic gas with
$c(\rho)= \rho^{(\ga-1)/2}$, an easy calculation gives
\begin{equation}\label{6-33}
  {\mathscr R}(r_1,r_2)=
\left(\frac{r_1}{r_2}\right)^{\beta},
  \quad\mbox{where}\quad \beta=\frac{3-\ga}{2(\ga-1)}.
\end{equation}
It is worth noticing that the approximation \eqref{Rapprox2} yields
the exact expression of the Riemann function for a classical
monoatomic gas with $\ga=5/3$ ($\beta=1$).  For other values of
$\beta$ the function $F$ in (\ref{5-29a}) can be shown to obey the
hypergeometric equation (see, e.g., Ref.~\cite{Som64}) and our
approximation corresponds to the first term in its series expansion.
Thus, we obtain
\begin{equation}\label{6-34}
\begin{split}
& W^{(1,3)}(r_+,r_-)\simeq \mp \left(\frac{2 }{r_+-r_-}\right)^{\beta}
\int_{-r_-}^{r_+}\! r^{\beta}\, \ox(r)\upd r,\\
&  W^{(2)}(r_+,r_-)\simeq \left(\frac2{r_+-r_-}\right)^{\beta} \\
& \phantom{W^{(2)}(r_+,r_-)} \times
  \left\{\int_{-r_-}^{r_m}\!\!\! r^{\beta}\, \ox(r)\upd r
+\int_{r_+}^{r_m}\!\!\! r^{\beta}\,  \ox(r)\upd r\right\}.
\end{split}
\end{equation}
For the case of ``shallow water'' equations with $\ga=2$ ($\beta=1/2$)
these formulae reproduce the results of
Refs.~\cite{ikp-19a,ikp-19b}. The approximation \eqref{6-34} cannot
be distinguished from the exact result of Riemann's approach for the
type of initial condition considered in Ref.~\cite{ikp-19a}.

We now study in some details a case where the dependence of $c$ on
$\rho$ is less simple than the one of Eq.~(\ref{1-2a}): this is the
case of a zero temperature Bose-Einstein condensate transversely
confined in an atomic wave guide. For a harmonic trapping, the
transverse averaged chemical potential can be represented by the
interpolating formula \cite{Ger04}
\begin{equation}\label{bec1}
\mu_\perp(\rho)=\hbar\omega_\perp\sqrt{1+4 a \rho},
\end{equation}
where $\omega_\perp$ is the angular frequency of the transverse
harmonic potential, $a>0$ is the $s$-wave scattering length, and
$\rho(x,t)$ is the linear density of the condensate. We note that
other expressions for $\mu_\perp$ have also been proposed in the
literature \cite{Sal02,Kam04}. Expression \eqref{bec1} yields the
correct sound velocity $m c^2=\rho\, \upd\mu_\perp/\upd\rho$ both in
the low ($a\rho\ll 1$) and in the high ($a\rho\gg 1$) density regimes.
In these two limiting cases the long wave length dynamics of the
system is thus correctly described by the hydrodynamic equations
(\ref{1-2a}) with, in appropriate dimensionless units:
\begin{equation}\label{bec2}
c^2(\rho)=\frac{\rho}{\sqrt{1+\rho}}\; ,
\end{equation}
where one has made the changes of variables $4 a\rho\to \rho$,
$u/u_0\to u$, $x/x_0\to x$ and $t/t_0\to t$, where $2 m
u_0^2=\hbar\omega_\perp$ and $t_0=x_0/u_0$. The length $x_0$ used to
non-dimensionalize the dispersionless equations is a free parameter:
we will chose it equal to the parameter $x_0$ appearing in the initial
condition \eqref{rho-init}. We note here that the initial condition
\eqref{rho-init} can be realized by several means in the context of
BEC physics. One can for instance suddenly
switch on at $t=0$ a blue detuned focused laser beam \cite{And98}.
An alternative method has been demonstrated in Ref.~\cite{Mat98}: by
monitoring the relative phase of a two species condensate, one can
implement a bump (or a through) in one of the components.

In the case characterized by Eqs.~\eqref{bec1} and \eqref{bec2},
expressions (\ref{Rapprox}) and \eqref{rderhobis} yield
\begin{equation}\label{bec4}
{\mathscr R}(r_1,r_2)
=\left(
\frac{\rho^2(r_1)+\rho^3(r_1)}{\rho^2(r_2)+\rho^3(r_2)}
\right)^{1/8},
\end{equation}
where $\rho(r)$ is the reciprocal function of
\begin{equation}\label{bec3}
r(\rho)=2\int_0^{\sqrt{\rho}}\frac{\upd u}{(1+u^2)^{1/4}}.
\end{equation}
In order to evaluate $W$ it then suffices to determine $\ox(r)$ by
inverting the relation \eqref{4-22} and to compute the appropriate
integrals \eqref{final}.  Once $W(r_+,r_-)$ is known in all three
regions 1, 2 and 3, it is possible to compute $r_+(x,t)$
and $r_-(x,t)$, and then $\rho(x,t)$ and $u(x,t)$ as explained in
Refs.~\cite{ikp-19a,ikp-19b}:

\begin{itemize}
\item[$\bullet$] One first determines the value $r^+_{1|2}(t)$
reached by $r^+$ at the boundary between regions 1 and 2, see
Fig.~\ref{fig1}(b). This boundary corresponds to the point where
$r^-=-r_m$ at time $t$. From Eqs.~\eqref{2-11}, $r^+_{1|2}(t)$ is thus
determined by solving
\begin{equation}
\frac{w_+^{(1)}(r^+_{1|2},-r_m)-w_-^{(1)}(r^+_{1|2},-r_m)}
{v_+(r^+_{1|2},-r_m) - v_-(r^+_{1|2},-r_m)} + t = 0\; ,
\end{equation}
where $w_+^{(1)}=\partial W^{(1)}/\partial r^+$.
We then know that, in region 1 at time $t$, $r^+$ takes all
possible values between $r_0$ and $r^+_{1|2}(t)$ (cf. Figs.~\ref{fig1}
and \ref{fig2}).

\item[$\bullet$] One then let $r^+$ vary in $[r_0,r_m]$. From 
Eqs.~\eqref{2-11}, at time $t$, the other Riemann invariant $r^-$ 
is solution of
\begin{equation}
\frac{w_+^{(1,2)}(r^+,r^-)-w_-^{(1,2)}(r^+,r^-)}
{v_+(r^+,r^-) - v_-(r^+,r^-)} + t = 0\; ,
\label{isoard_eq_dispn}
\end{equation}
where the superscript should be (1) if $r^+ \in [ r_0,
  r^+_{1|2}(t)]$ and (2) if $r^+ \in [r^+_{1|2}(t),r_m]$.

\item[$\bullet$] At this point, for each value of $t$ and $r^+$ we
  have determined the value of $r^-$. The position $x$ is then
  obtained by either one of Eqs.~\eqref{2-11}. So, for given $t$ and
  $r^+$ in regions 1 and 2, one has determined the values
  of $r^-$ and $x$.  In region 3 we use the symmetry of the
  problem and write $r^\pm(x,t) = -r^\mp(-x,t)$.
\end{itemize}

\begin{figure}
\begin{center}
\includegraphics[width=\linewidth]{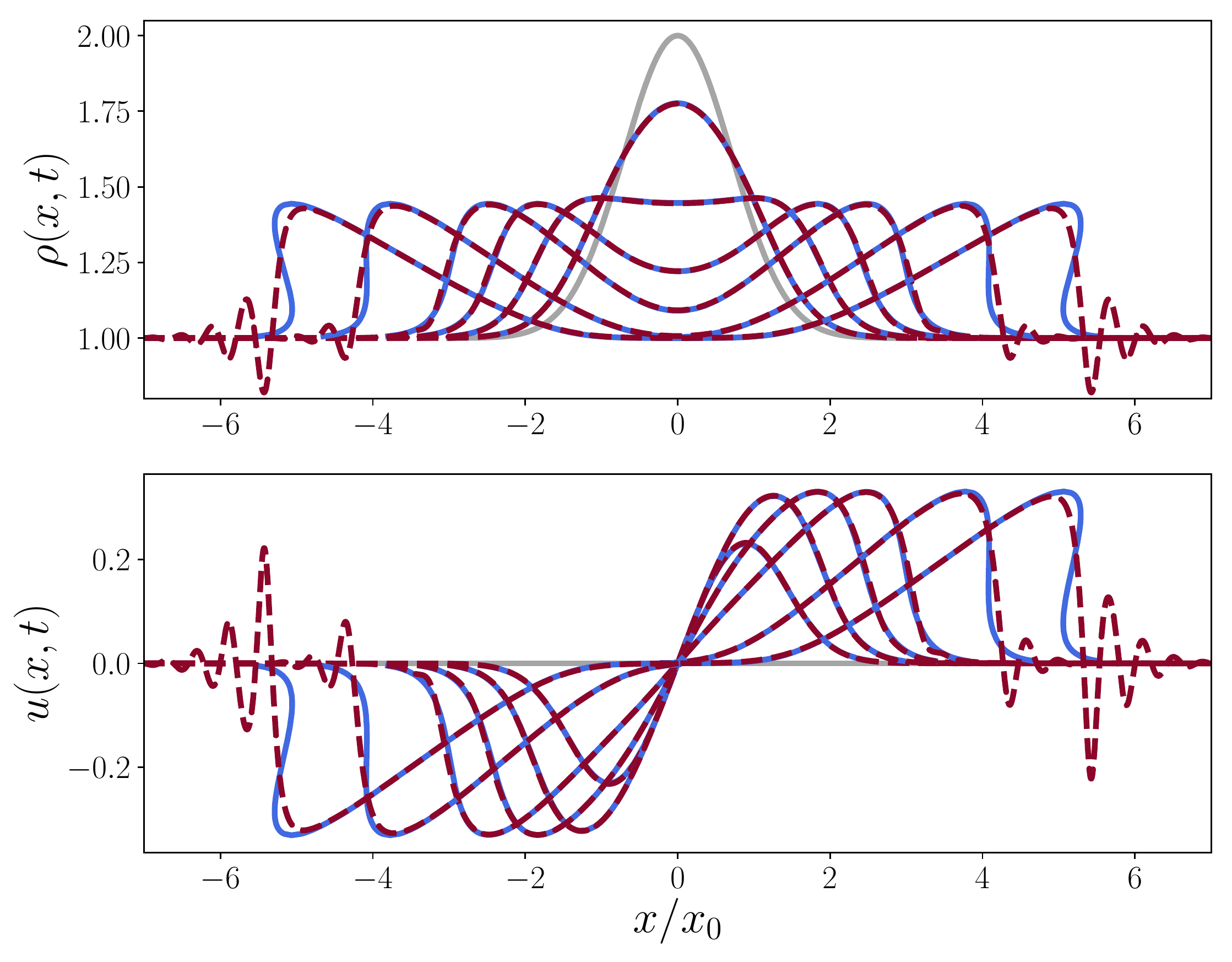}
\caption{Density and velocity plotted as a function of $x/x_0$ for
  dimensionless times $t/t_0=0.5$, 1, 1.5, 2, 3 and 4 respectively.
  The initial conditions are given by Eqs.~\eqref{1-3} and
  \eqref{rho-init} with $\rho_0=1$ and $\rho_1=1$, they are represented
  by the gray solid lines. The blue solid lines are the results of the
  hydrodynamic system \eqref{1-1} obtained from the approximate
  Riemann's approach described in the text. The dashed lines are the
  results of the numerical simulations of Eq.~\eqref{bec5}.}
\label{fig3}
\end{center}
\end{figure}

The above procedure defines a mapping of the whole physical
$(x,t)$ space onto the hodograph $(r^+,r^-)$ space.  The density and
velocity profiles are then obtained by means of Eqs.~\eqref{2-5}. The
results are compared with numerical simulations in Fig.~\ref{fig3} for
an initial profile \eqref{rho-init} with $\rho_0=1$ and
$\rho_1=1$. The simulations have been performed by solving numerically
a generalized nonlinear Schr\"odinger equation of the form
\begin{equation}\label{bec5}
{\rm i} \psi_t=-\tfrac12 \psi_{xx} + 2\,  \psi \sqrt{1+\rho},
\end{equation}
where $\rho(x,t)=|\psi|^2$, $u(x,t)=(\psi^*\psi_x-\psi\, \psi_x^*)/(2
{\rm i} \rho)$ and $\psi(x,0)=\sqrt\rho(x,0)$. This effective
Gross-Pitaevskii equation reduces to the system \eqref{1-1} with the
speed of sound \eqref{bec2} in the dispersionless limit\footnote{It
  would be easier and more natural to compare our approximate Riemann
  approach with the numerical solution of Eqs.~\eqref{1-1}. However,
  the difference between the two results is so small that the
  discussion of this comparison has little interest.}. It yields an
excitation spectrum always of Bogoliubov type, which is incorrect in
the large density limit ($\rho\gg 1$).  However, one can show that
Eq.~\eqref{bec5} is acceptable even in this limit provided one remains
in the long wave-length, hydrodynamic regime. It is not appropriate
when rapid oscillations appear in the density and velocity (if
$\rho\gg 1$) such as observed in Fig.~\ref{fig3} for $t/t_0=3$. These
oscillations correspond to the onset of a dispersive shock wave, which
occurs at a time denoted as the wave breaking time: $t_{\rm\sss
  WB}$. For $t>t_{\rm\sss WB}$ the numerical simulations can be
considered as accurately describing the physical system only when
$\rho\ll 1$. But for $t>t_{\rm\sss WB}$ our dispersionless approach
also fails (see below): we are thus safe when comparing our results
with numerical simulations at earlier times.

One sees in Fig.~\ref{fig3} that our solution of the hydrodynamic
equations \eqref{1-1} agrees very well with the numerical simulations
of the dispersive equation \eqref{bec5} at short time. For larger
times the profile steepens, eventually reaching a point of gradient
catastrophe at time $t_{\rm\sss WB}$. It is thus expected that for
$t\simeq t_{\rm\sss WB}$ the solution of the dispersionless system
\eqref{1-1} departs from the numerical simulations, as seen in the
figure. However, this difference is not a sign of a failure of our
approximation, but it rather points to the breakdown of the
hydrodynamic model \eqref{1-1}. After $t_{\rm\sss WB}$ the system
\eqref{1-1} leads to a multi-valued solution if not corrected to
account for dispersive effects, as can be seen in Fig.~\ref{fig3}.

\section{Wave breaking time}

We now turn to the determination of the wave breaking time
$t_{\rm\sss WB}$ at which a shock is formed. After $t_{\sss\rm WB}$
the system \eqref{1-1} has to be modified in order to account for
viscous and/or dispersive effects, depending on the physical situation
under consideration.

We treat the case of an initial profile roughly
of the type \eqref{rho-init}: a bump over a uniform
background. Wave breaking corresponds to the occurrence of a gradient
catastrophe for which $\partial r_{\pm}/\partial x=\infty$.  If one
considers for instance the right part of the profile (region 3),
from Eq.~\eqref{2-11}, this occurs at a time $t$ such that
\begin{equation}\label{wb1}
t=-\frac{\partial w_+^{(3)}/\partial r_+}{1 + c'(r_+-r_-)}
=-\frac{\partial w^{(3)}_+/\partial r_+}
{1+\left.\frac{\displaystyle \upd \ln c}
{\displaystyle\upd \ln\rho}\right|_{r_+-r_-}},
\end{equation}
and $t_{\rm\sss WB}$ is the smallest of the times \eqref{wb1}.
It is worth noticing that this formula yields an expression
for the breaking time obtained from our approximate solution 
of the initial value problem and in this sense it provides
less general but more definite result than the upper estimate 
of the breaking time obtained by Lax in Ref.~\cite{Lax64}.

One can easily compute $t_{\rm\sss WB}$ approximately when the point
of largest gradient in $\overline{\rho}(x)$ lies in a region where
$\overline{\rho}\simeq \rho_0$. This occurs for some specific initial
distributions (such as the inverted parabola considered in
Ref. \cite{ikp-19a}) or when the initial bump is only a small
perturbation of the background. In this case, it is legitimate to
assume that wave breaking is reached for $r_-\simeq -r_0$ and that
\begin{equation}
  r_+-r_-\simeq \tfrac12 \int_0^\rho \frac{c(\rho')}{\rho'}\upd\rho'
  +\tfrac12 \int_0^{\rho_0} \frac{c(\rho')}{\rho'}\upd\rho'.
\end{equation}
Eqs.~\eqref{3-15} and \eqref{final} then lead to
$w_+^{(3)}\simeq \ox(r_+)$ and \eqref{wb1} becomes
\begin{equation}\label{wb2}
t\simeq -\frac{2}{1+\left.\frac{\displaystyle \upd \ln c}
{\displaystyle\upd \ln\rho}\right|_{\overline{\rho}}}
\times
\frac{\overline{\rho}}
{c(\overline{\rho})\; \frac{\displaystyle \upd \overline{\rho}}
{\displaystyle\upd x}}\; ,
\end{equation}
where $\overline{\rho}$ stands for $\overline\rho(\ox(r^+))$.  Within
our hypothesis, it is legitimate to assume that the shortest of times
$t$ is reached close to the point $\ox(r^+)$ for which $|\upd
\overline{\rho}/\upd x|$ is maximal. We note $x^*$ the
coordinate of this point and $\rho^*=\overline{\rho}(x^*)$. One thus
obtains
\begin{equation}\label{wb3}
t_{\rm\sss WB}\simeq
\frac{2}{1+\left.\frac{\displaystyle \upd \ln c}
{\displaystyle\upd \ln\rho}\right|_{\rho^*}}
\times
\frac{\rho^*}{c(\rho^*)\,\cdot\,{\rm max}
\left|\frac{\displaystyle \upd \overline{\rho}}
{\displaystyle \upd x}\right|}\; .
\end{equation}
In a ``shallow water'' case with $\gamma=1/2$ and for an initial profile
where the bump is an inverted parabola, such as considered in
Ref.~\cite{ikp-19a}, the above formula is exact.

For the initial profile \eqref{rho-init}, in the case where the speed
of sound is given by \eqref{bec2}, formula \eqref{wb3} yields
\begin{equation}\label{wb4}
  t_{\rm\sss WB}\simeq \sqrt{\frac{e}{2}}\, \frac{8(1+\rho^*)}{6+5\rho^*}
  \frac{x_0}{c(\rho^*)} \frac{\rho^*}{\rho_1}.
\end{equation}
The location $x_{\sss\rm WB}$ of the wave breaking event can be obtained
from \eqref{2-11}. Within our approximation scheme, this yields, for
the right part of the profile:
\begin{equation}\label{wb5}
  x_{\sss\rm WB}\simeq x^*+c(\rho^*) t_{\sss\rm WB}.
\end{equation}

\begin{figure}
\begin{center}
\includegraphics[width=\linewidth]{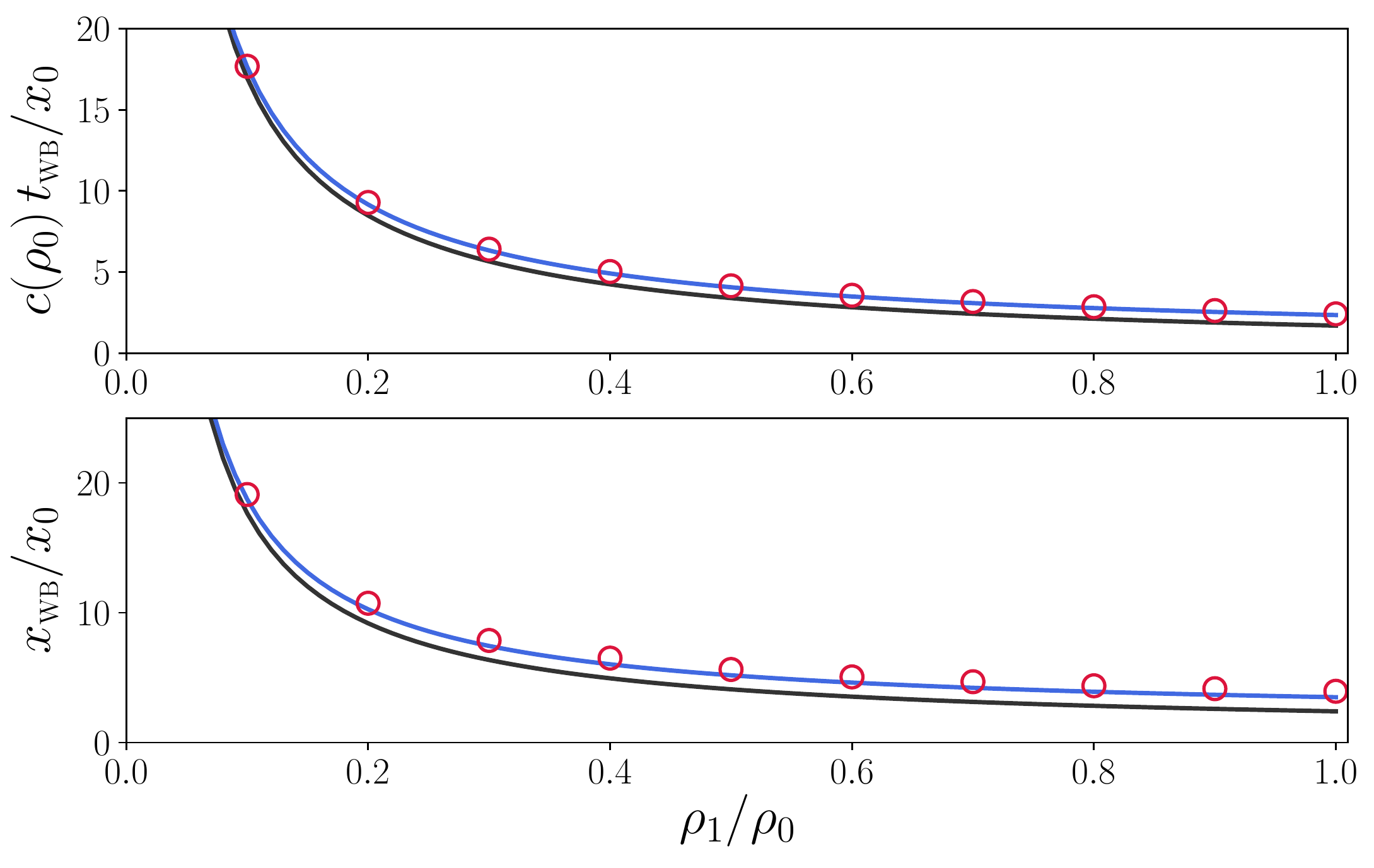}
\caption{Wave breaking time $t_{\sss\rm WB}$ and position of the wave
  breaking event $x_{\sss\rm WB}$ for different values of
  $\rho_1/\rho_0$. The system considered is a quasi-1D BEC for which
  the speed of sound in given by \eqref{bec2}. The initial profile is
  given by Eqs.~\eqref{1-3} and \eqref{rho-init}.  The blue solid
  lines are the approximate results \eqref{wb4} and \eqref{wb5}. The
  red dots are the results obtained from Riemann's approach. The black
  solid lines are obtained by replacing $\rho^*$ by $\rho_0$ in
  Eqs.~\eqref{wb4} and \eqref{wb5}, see the text.}
\label{fig4}
\end{center}
\end{figure}
These results are compared in Fig.~\ref{fig4} with the values
determined from the Riemann approach. The overall agreement is
excellent. We also note that replacing $\rho^*$ by $\rho_0$ in
Eqs.~\eqref{wb4} and \eqref{wb5} gives a result which is less
accurate, but still quite reasonable, see Fig.~\ref{fig4}.

\section{Conclusion}

We have presented an approximate method for describing the
hydrodynamic evolution of a nonlinear pulse. The method is quite
general and applies for any type of nonlinearity. It has been tested
for cases of experimental interest in the context of nonlinear optics
in Ref.~\cite{ikp-19a} and here for studying the spreading of a
nonlinear pulse in a guided atomic Bose-Einstein condensate. This last
example is of particular interest for bench-marking the approach
because the nonlinearity at hand has a non-trivial density dependence.

One could imagine to extend the present study in several
  directions. A possible track would be to solve the dispersionless
  shallow water equations [Eqs. \eqref{1-1} and \eqref{1-2a} with
    $\gamma=2$] for more general initial conditions than discussed in
  the present work, as considered for instance in
  Refs. \cite{Pel13,Rod19} in the context of the initial stage of
  formation of a tsunami. Future studies could also test the present
approach in the optical context for pulses propagating in a nonlinear
photo-refractive material, where, up to now, no theoretical method was
known for dealing with the dispersionless stage of evolution.  In this
context we note that the simple and accurate approximate analytic
results obtained for $t_{\sss\rm WB}$ and $x_{\sss\rm WB}$
[Eqs.~\eqref{wb4} and \eqref{wb5}] should be helpful for determining
the best parameters for an experimental observation of the wave
breaking phenomenon.

We finally stress that the approximate scheme presented in this
work, providing an accurate account of the stage of non-dispersive
propagation of a pulse, is an important and necessary step for
studying the post wave breaking dynamics, and particularly the
formation of dispersive shock waves in non-integrable systems.

\end{document}